\documentclass[12pt,reqno]{amsart}
 \usepackage{amsmath,amssymb}
\usepackage{geometry}                		
\usepackage{graphicx}
\usepackage{multirow}

\newcommand{\rb}{\mathbf{r}}
\newcommand{\Rb}{\mathbf{R}}
\newcommand{\nb}{\hat{\mathbf{n}}}
\newcommand{\cm}{\text{\ cm}}

\title{Learning the detector in optical tomography}
\author{Zijian Wang}
\address{Department of Applied and Computational Mathematics, Yale University, New Haven, CT, USA}
\email{zijian.wang@yale.edu}
\author{Andreas Hauptmann}
\address{Research Unit of Mathematical Sciences, University of Oulu, Oulu, Finland \\
Department of Computer Science, University College London, London, UK}
\email{Andreas.Hauptmann@oulu.fi}
\author{Lu Lu}
\address{Department of Statistics and Data Science, Yale University, New Haven, CT, USA}
\email{lu.lu@yale.edu}
\author{John C. Schotland}
\address{Department of Mathematics and Department of Physics, Yale University, New Haven, CT, USA}
\email{john.schotland@yale.edu}

\begin{document}

\begin{abstract}
We propose a method to reconstruct the optical absorption of a highly-scattering medium probed by diffuse light. The method consists of learning the optical detection system and then using this result to reconstruct the absorption. Our results are illustrated by numerical simulations.
\end{abstract}

\maketitle

The development of methods for imaging of highly-scattering media, such as clouds, colloidal suspensions and biological tissue, is a topic of long-standing interest~\cite{arridge1999optical}. Optical tomography  
is one such method that has been widely employed in biomedical imaging. 
In a typical experiment, a sample is illuminated by a narrow collimated beam. The transmitted or reflected light is collected by an array of detectors, often a CCD camera, 
while the beam is scanned over the surface of the medium. 
The inverse problem of optical tomography is to reconstruct the optical properties of the medium from such measurements. A variety of approaches to this problem have been investigated~\cite{arridge2009optical}. These include optimization methods, often formulated in a Bayesian framework, and direct inversion methods, such as the inverse Born series. More recently, machine learning algorithms have also been employed, either as a component of an optimization method or as a stand-alone algorithm \cite{chen2020physics,arridge2019solving,fan2020solving,karniadakis2021physics,zhu2023fourier}, with applications to optical tomography \cite{mozumder2021model,yoo2019deep,kamilov2015learning,mozumder2024diffuse}. 

Regardless of the method of reconstruction, a crucial point should be emphasized. The image that is recorded by the camera does not directly correspond to the \emph{ideal data} that is taken as the input to a reconstruction algorithm. More precisely, consider the setting in which light propagation is described by radiative transport theory~\cite{carminati2021principles}. The specific intensity, measured at a point on the surface of the medium in a fixed outward direction, corresponds to the ideal data. However,  the camera does not measure this quantity. Rather, the camera records a geometrically weighted average of the specific intensity over directions that pass through its aperture. In practice, this problem is addressed by calibration of the experimental apparatus~\cite{wang2005experimental,konecky2008imaging}. By focusing the camera at infinity and collecting the quasiparallel rays that exit the medium over a small range of directions, the camera measurements are taken as an approximation to the ideal data. 

In this Letter, we develop a method to recover ideal data from camera measurements. The method is based on the principles of geometrical optics, in which the propagation of light from the surface of the medium to the camera is directly accounted for. This approach has several advantages over existing methods. It is based on a first-principles 
theory rather than an empirical procedure, and is analyzable and systematically improvable. 
More practically, the measured light is collected over a large angular range of directions, which reduces the overall data-collection time of the experiment.
Finally, the method can be integrated into a two-step reconstruction algorithm. In the first step, the ideal data is recovered using the ray-optics model. This entails learning the mapping from camera data to ideal data by minimizing a suitable objective function.
The second step consists of recovering the optical absorption of the medium from the ideal data. Here we utilize either a standard regularized inversion method or an encoder-decoder~\cite{hinton2006reducing,cho2014learning} convolutional neural network (CNN)~\cite{lecun2002gradient,krizhevsky2012imagenet}. Alternatively, we consider an end-to-end inversion procedure using a CNN to recover the absorption from camera data. The performance of all three algorithms is compared with numerical simulations.

\begin{figure}[t]
\centering
\includegraphics[width=0.6\linewidth]{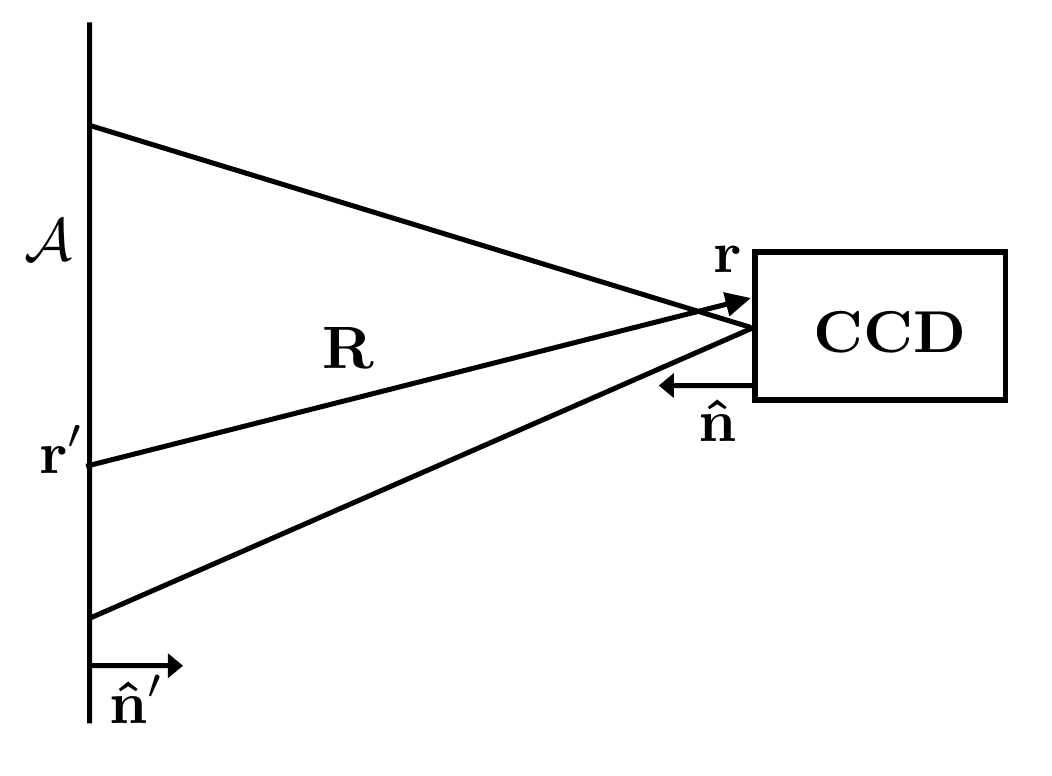}
\caption{Illustrating the experimental geometry.}
\label{fig:camera}
\end{figure}

We begin by considering the propagation of diffuse light in an absorbing medium. 
The density of electromagnetic energy $u$ obeys the diffusion equation
\begin{align}
\label{eqn: light model}
 -D\nabla^2 u + \sigma(\rb) u  &= 0 \quad \text{ in } \quad \Omega , \\
 u&=g \quad \text{ on } \quad \partial\Omega ,
\end{align}
where $\Omega$ is a three-dimensional volume, $\sigma$ is the absorption coefficient, $D$ is the diffusion constant, and $g$ is the source~\cite{carminati2021principles}. 
Following standard procedures~\cite{arridge2009optical}, we find that within the accuracy of the Born approximation, the solution to \eqref{eqn: light model} is of the form
\begin{align}
\label{eqn: integral equation}
u(\rb) =u_0(\rb)+\int_\Omega G(\rb,\rb')\delta\sigma(\rb')u_0(\rb')d\rb'.
\end{align}
Here the incident field $u_0$ is the solution to \eqref{eqn: light model} in a reference medium with constant absorption $\sigma_0$, $G$ is the Green's function for the diffusion equation with $\sigma=\sigma_0$, and $\delta\sigma(\rb)=\sigma(\rb)-\sigma_a$.
The specific intensity $I(\rb,\hat{\mathbf{k}})$ of light that exits the medium
at the point  $\rb\in\partial\Omega$ in the direction $\hat{\mathbf{k}}$ is given by 
\begin{align}
\label{eqn: specific intensity}
    I(\rb,\hat{\mathbf{k}}) = -D\hat{\mathbf{k}}\cdot\nabla u(\rb) .
\end{align}
It follows that the relative intensity at a point on the boundary in the outward normal direction $\nb$ is given by
\begin{align}
\label{eqn: born normal}
\Delta I(\rb) = \int_\Omega  \frac{\partial}{\partial \nb(\rb)} G(\rb,\rb') \delta\sigma (\rb')u_0(\rb')d\rb' ,
\end{align}
where $\Delta I(\rb) = I(\rb,\nb)-I_0(\rb,\nb)$, $I_0$ is the specific intensity in the reference medium and $\rb\in\partial\Omega$. The light exiting the medium is imaged by an optical system onto the camera, as shown in Fig.~\ref{fig:camera}. 

The power measured by the camera at the point $\rb$ is given by
\begin{align}
\label{eq:power}
P(\rb) = \int_{\mathcal A} K(\rb,\rb')I(\rb',\hat{\Rb})d\rb' ,
\end{align}
where $\Rb=\rb-\rb'$ and $\mathcal A$ is the portion of $\partial\Omega$ that is visible to the camera. Here $K$ accounts for the local orientation of the transmitted ray relative to $\partial\Omega$ and the camera, and the $1/R^2$ loss of intensity with propagation:
\begin{align}
\label{eq:K}
K(\rb, \rb') = \Delta A f(\nb'\cdot\hat \Rb)\frac{(\nb\cdot\hat \Rb)  (\nb'\cdot\hat \Rb)}{|\Rb|^2} ,
\end{align}
where $\nb$ is the normal to $\partial\Omega$, $\nb'$ is the normal to the plane of the camera, $\Delta A$ is the detector area, and the function $f$ is determined by the details of the optical system. For simplicity, we will take $f$ to be a Gaussian function of the form
$f(x)=\exp(-x^2/\sigma^2)$, where $\sigma$ is constant. 

Using the above model, the forward problem consists of evaluating the map from the absorption to detector measurements. The map is evaluated by first solving the diffusion equation \eqref{eqn: light model} to obtain the specific intensity and then computing the power from \eqref{eq:power}. The diffusion equation is solved using the finite-element method implemented in DOLFINx. The boundary sources are taken to be of the form $g(\rb)=\exp(-|\rb-\rb_0|^2/\Delta^2)$, where $\rb_0\in\partial\Omega$ and  $\Delta$ is a constant. Adaptive mesh refinement is employed around the source, resulting in a precision of $7$-digits. 

The inverse problem is solved by a two-step reconstruction method. The first step consists of recovering the ideal data from camera measurements using a data-driven approach. We denote by $A: x\mapsto y$
the map from camera data $x$ to ideal data $y$. Here $x, y \in \mathbb R^{N_1\times N_2}$ correspond to measurements of $P$ and $I$, respectively, sampled on a two-dimensional uniform grid of size $N_1\times N_2$,
and $A$ is a $N_1N_2\times N_1N_2$ matrix.
Given the training data $(x_n, y_n)$, where $n=1,\ldots, N$, the map $A$ is learned
by minimizing the regularized mean-squared error
\begin{align}
\label{objective_function}
\mathcal L  = \frac{1}{N}\sum_{n=1}^{N} |Ax_n-y_n|^2
    + \frac{\lambda}{2}\text{Tr}(A^TA) ,
\end{align}
where $\lambda$ is the regularization parameter. The second term in $\mathcal L$ is introduced to compensate for the low-rank nature of $A$. We employed a singular value decomposition based solver to compute the minimizer of \eqref{objective_function}. To select the regularization parameter $\lambda$, the dataset $(x_n,y_n)$ was randomly partitioned into a training set (80\% of the samples) and a validation set (20\% of the samples). For each candidate value of $\lambda$ in a prescribed range, the mapping $A$ was trained on the training subset, and the corresponding validation error was computed on the withheld subset. The optimal $\lambda$ was then chosen as the value that minimized the mean-squared prediction error on the validation set.
The data $(x_n,y_n)$ is obtained by solving the diffusion equation \eqref{eqn: light model} for a set of absorptions $\delta\sigma_n$ and then computing $y_n$ from \eqref{eq:power}. 
Once $A$ has been determined, the ideal data $y$ is recovered from camera measurements $x$ using the relation $y=Ax$.
\begin{figure}[t]
\centering
\includegraphics[width=.8\linewidth]{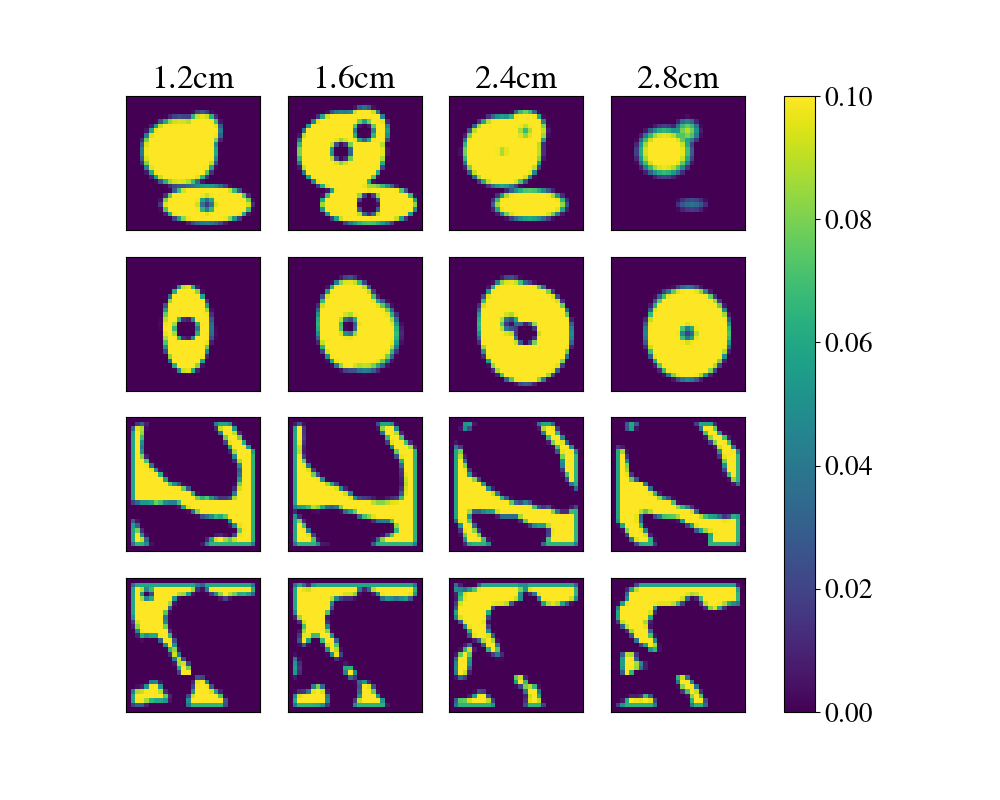}
\caption{Comparison of in-distribution ellipsoid (top two rows) and out-of-distribution breast (bottom two rows) absorptions, with each column showing a cross-section of the three-dimensional absorption image at the indicated height.}
\label{fig:sigma_example_3d}
\end{figure}
\begin{figure}[h]
\centering
\includegraphics[width=0.8\linewidth]{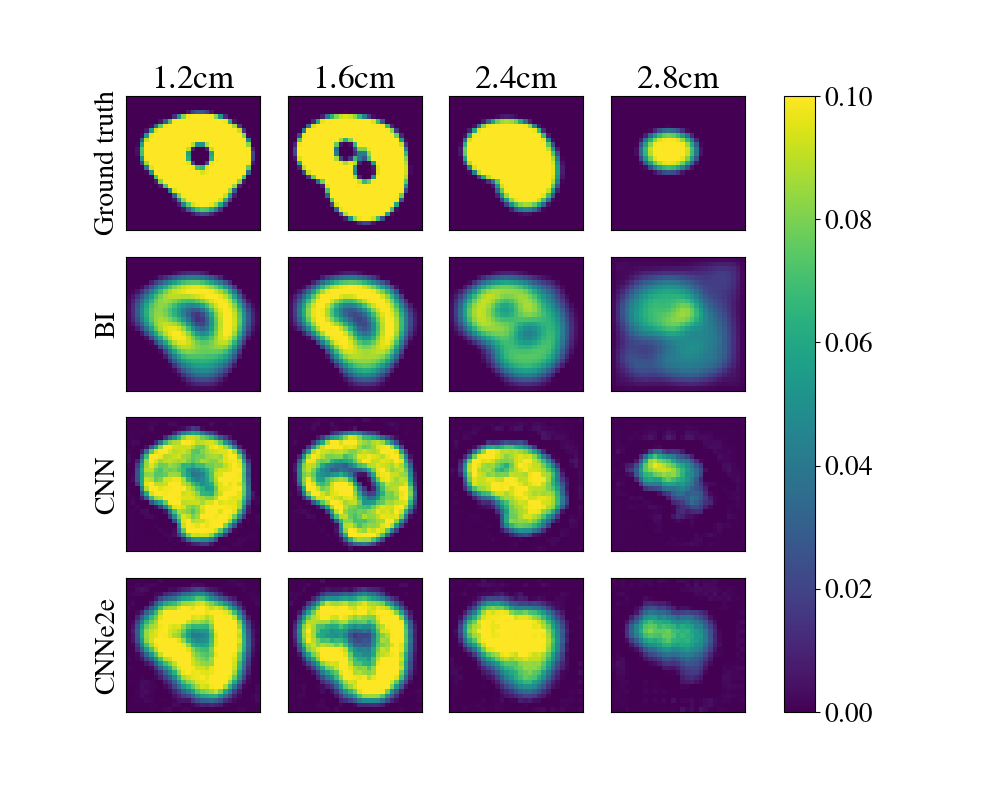}
\caption{Reconstruction examples for samples from $\mathcal{D}_{\text{ellipsoid}}$, with rows showing the ground truth, Born inversion (BI), CNN, and end-to-end CNN (CNNe2e) results, and columns corresponding to slices at the indicated heights.}
\label{fig:in_distribution3d}
\end{figure}
\begin{figure}[h]
\centering
\includegraphics[width=0.8\linewidth]{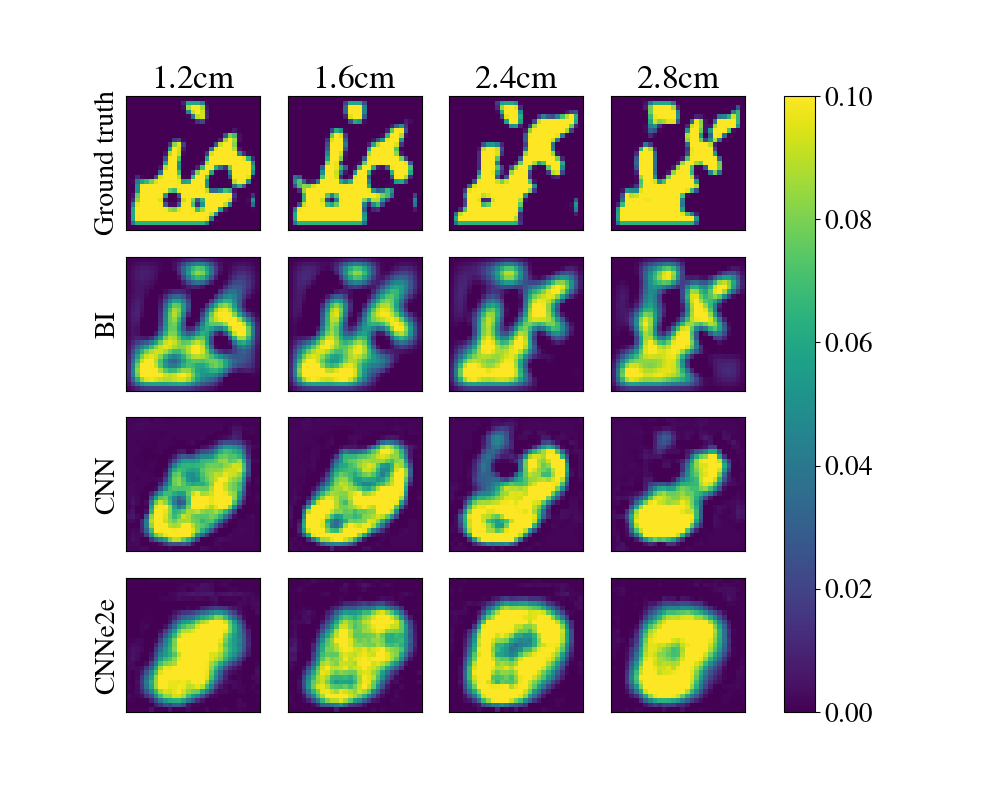}
\caption{Same as Fig. \ref{fig:in_distribution3d}, but with breast data.}
\label{fig:out_of_distribution3d}
\end{figure}

The second step consists of recovering $\delta\sigma$ from the ideal data $y$. We consider two approaches to this problem: numerical inversion of the integral equation \eqref{eqn: born normal} and deep learning. The former solves the system of linear equations that arises from discretizing 
\eqref{eqn: born normal}. We will refer to this method as Born inversion; it is carried out as previously described \cite{markel2019fast}.
The latter approach is data-driven and makes use of  a dataset $\mathcal{D}$ consisting of pairs of ideal measurements and discretized absorption coefficients $\delta\sigma$. To proceed, we
consider $M_1\times M_2$ sources on $\partial\Omega$.
We also discretize $\delta\sigma$ on a grid, where $\delta\sigma\in\mathbb{R}^{P_1\times P_2\times P_3}$.
Let $f_\Theta:\mathbb{R}^{M_1\times M_2\times N_1\times N_2}\to \mathbb{R}^{P_1\times P_2\times P_3}$ be a map from the space of measurements to the space of absorptions, represented by a neural network parametrized by $\Theta$.
We use an encoder-decoder architecture, so that $f_\Theta:=g_{\Psi}\circ h_{\Phi}$, where $h_{\Phi}:\mathbb{R}^{M_1\times M_2\times N_1\times N_2}\to\mathbb{R}^{d}$ and $g_{\Psi}:\mathbb{R}^{d}\to\mathbb{R}^{P_1\times P_2\times P_3}$ represents the encoder and decoder maps of the high-dimensional measurement data into and from a latent space of dimension $d$.
This architecture is a generalization of a rate-distortion auto-encoder for learning low-dimensional representations \cite{balle2016end}.
We construct both the encoder $h_\Phi$ and the decoder $g_\Psi$ using convolutional layers.
A convolutional layer is a spatially invariant operation that performs a convolution over a small grid centered at each pixel, a common building block in deep learning \cite{lecun2015deep}.
Unlike conventional two-dimensional convolution, where each pixel is real-valued, we allow pixels to be vector-valued.
The dimension of each pixel vector is known as the number of channels. That is, the encoder $h_\Phi$ takes inputs of spatial dimension $N_1\times N_2$ with $M_1\times M_2$ channels per pixel. We note that our choice of architecture is commonly used in machine learning applications for imaging \cite{ronneberger2015u}.
The neural network $f_\Theta$ is trained by minimizing the error between the predicted and true absorptions. Given training data $(x_n,y_n)$, where $x_n$ denotes the measurement data and $y_n$ the corresponding absorption, the error function is of the form
\begin{align}
\mathcal{E} = \frac{1}{N}\sum_{n=1}^N \bigl\| f_\Theta(x_n) - y_n \bigr\|^2 + \frac{\lambda}{2}\|\Theta\|^2 ,
\end{align}
where $\|\cdot\|$ denotes the Euclidean norm. The first term penalizes discrepancies between  predicted absorptions and true absorptions, while the second term penalizes the growth of the network weights \cite{zhang2018three}. This procedure is similar to the linear case in \eqref{objective_function}, except that the matrix $A$ is replaced by the neural network $f_\Theta$ and the Frobenius norm $\text{Tr}(A^TA)$ is replaced with the term $\|\Theta\|^2$.  
The network was trained by using the Adam optimizer \cite{kingma2014adam}.

We now illustrate the above reconstruction algorithms with numerical simulations. 
We suppose that $\Omega$ is a rectangular prism of dimensions $6\cm\times 6\cm\times 4\cm$.
Light is incident on the bottom rectangle and is collected on the top rectangle of the medium.
The voxel size and grid spacing are taken to be $0.2\cm$ both for the discretization of the measurements and the absorption coefficient $\delta\sigma$. Thus $M_1=M_2=29$, $N_1=N_2=29$ and $P_1=P_2=29, P_3=19$. The camera measurements are subject to shot noise, which scales with the square root of the intensity. Following \cite{konecky2008imaging}, we add Gaussian noise to all pixels with a standard deviation of 1\%  relative to the detector measurements, for all pixels.
We take the background absorption $\sigma_0 = 1.0 \cm^{-2}$ \cite{units} and consider two datasets corresponding to different choices of the absorption $\delta\sigma$. The first dataset  
$\mathcal{D}_{\text{ellipsoid}}$ consists of unions of ellipsoids with spherical inclusions, as illustrated in Fig.~\ref{fig:sigma_example_3d}.
The second dataset $\mathcal{D}_{\text{breast}}$ contains segments of breast magnetic resonance images \cite{lou2017generation}. In both datasets, we vary the support of $\delta\sigma$ and set its value to be $0.1 \cm^{-2}$.
See Fig.~\ref{fig:sigma_example_3d} for a comparison of the datasets. We use $70\%$ of $\mathcal{D}_{\text{ellipsoid}}$ for training, $20\%$ of $\mathcal{D}_{\text{ellipsoid}}$ for early stopping, and the remaining $10\%$ for out-of-sample testing. The dataset $\mathcal{D}_{\text{breast}}$ is used for out-of-distribution testing. To evaluate the performance of the reconstruction algorithms, we apply three different metrics: relative $l_1$ error, relative $l_2$ error and the structural similarity index (SSIM) \cite{wang2004image}.
In addition to these reconstruction algorithms, we have also trained an end-to-end CNN encoder-decoder (CNNe2e) network that directly maps camera measurements to $\delta\sigma$. 
The CNNe2e network makes use of the same encoder-decoder architecture as the CNN.
Thus, the optical system is learned along with the inversion procedure.
Table \ref{tab:benchmark_full} shows the comparison between CNN, CNNe2e and Born inversion for all three metrics. Reconstructions using ideal data, camera data and recovered data are shown. 
Within the distribution $\mathcal{D}_{\text{ellipsoid}}$, we observe that CNN outperforms Born inversion. However, Born inversion significantly outperforms CNN on out-of-distribution data $\mathcal{D}_{\text{breast}}$. 
In both cases, the end-to-end CNN algorithm does not perform as well as the CNN.
This finding is consistent with the expectation that neural networks learn in-distribution examples, but typically fail to extrapolate to out-of-distribution examples \cite{novak2018sensitivity,zhu2023reliable}.
Figs.~\ref{fig:in_distribution3d} and \ref{fig:out_of_distribution3d} further illustrate the comparison of in-distribution and out-of-distribution performance for Born inversion, CNN and CNN2e. 
We see that CNN slightly outperforms Born inversion for predicting in-distribution ellipsoid phantoms. Both Born inversion and CNN reconstructions show the presence of inclusions, but neither accurately reconstructs them. However, both the Born inversion and CNN methods are able to reconstruct the boundaries of the reconstructed objects.
In addition, we find that CNNe2e does not perform as well as CNN. Moreover, as shown in 
Fig.~\ref{fig:out_of_distribution3d}, deep learning methods generally fail to reconstruct out-of-distribution data. As expected, the performance of Born inversion is similar to the in-distribution test cases in Fig.~\ref{fig:in_distribution3d}. 
The out-of-distribution tests for the deep learning methods (rows $3,4$ in Fig.~\ref{fig:out_of_distribution3d}) produce ellipsoid-like reconstructions, similar to the training data, even when the true targets are breast images. This is a consequence of the data-driven nature of the method.

\begin{table}[h]
\centering
\begin{tabular}{llllll}
\hline
\textbf{Dataset} & \textbf{Input} & \textbf{Method} & \textbf{$l_2$ error} & \textbf{$l_1$ error} & \textbf{SSIM} \\ \hline
\multirow{6}{*}{Ellipsoid} 
    & \multirow{2}{*}{Ideal}     & BI & $0.467$ & $0.743$ & $0.554$ \\
    &                            & {CNN}  & $0.344$ & $0.330$ & $0.794$ \\ \cline{2-6}
    & \multirow{2}{*}{Recovered} & BI & $0.468$ & $0.744$ & $0.554$ \\
    &                            & {CNN} & $0.351$ & $0.372$ & $0.780$ \\ \cline{2-6}
    & \multirow{2}{*}{Camera}  & BI & $94.696$ & $160.748$ & $0.000$ \\
    &                            & {CNNe2e}  & $0.424$ & $0.465$ & $0.666$ \\ \hline
\multirow{6}{*}{Breast} 
    & \multirow{2}{*}{Ideal}     & {BI} & $0.439$ & $0.667$ & $0.559$ \\
    &                            & CNN  & $0.917$ & $1.050$ & $0.149$ \\ \cline{2-6}
    & \multirow{2}{*}{Recovered} & {BI} & $0.440$ & $0.669$ & $0.559$ \\
    &                            & CNN & $0.913$ & $1.067$ & $0.159$ \\ \cline{2-6}
    & \multirow{2}{*}{Camera}  & BI & $95.031$ & $154.944$ & $0.000$ \\
    &                            & {CNNe2e}  & $0.996$ & $1.228$ & $0.102$ \\ \hline
\end{tabular}
\caption{Comparison of the performance of Born inversion (BI), convolutional neural network (CNN) and end-to-end CNN (CNNe2e) reconstruction algorithms. Results for the error metrics $l_2$-error, $l_1$-error and structure similarity (SSIM) are shown. 
Note that higher values of SSIM indicate smaller errors.
The three categories are reconstructions from ideal data (computed from \eqref{eqn: born normal}), from camera data (computed from \eqref{eq:power}) and from recovered data.}
\label{tab:benchmark_full}
\end{table}

In conclusion, we have proposed an image reconstruction method for optical tomography with diffuse light. The method consists of two steps. First, the ideal data is recovered using a first-principles model that learns the mapping from camera data to ideal data. The second step consists of recovering the optical absorption of the medium from the ideal data, and is carried out
using either Born inversion or an encoder-decoder CNN. We also consider an end-to-end inversion procedure using a CNN to recover the absorption directly from camera data. In general, we find that deep learning performs somewhat better than Born inversion for in-distribution tests, but fails for out-of-distribution tests. This is not unexpected since deep learning is data-driven, while Born inversion is not, consistent with the principle \emph{only learn what cannot be modeled}.  In future work, we plan to incorporate a nonlinear inversion method, such as the inverse Born series, in the second step of the image reconstruction algorithm. This would afford the possibility of reconstructing images with higher absorption contrast than can be obtained using Born inversion. Finally, the problem of learning an optical system from measurements is potentially of interest beyond the setting of optical tomography. For instance, within radiometry, Eqs. (\ref{eq:power}) and (\ref{eq:K}) describe a general relation between an incident specific intensity and the output of the system, as characterized by the function $f$. We note that learning $f$, rather than the kernel $K$, has some advantages since the latter problem is relatively ill-posed due to the low rank nature of $K$.

\bibliographystyle{plain}
\bibliography{sample}

\begin{thebibliography}{10}
\bibitem{units}
We work in units where the diffusion coefficient $D=1$. This means that the
absorption $\sigma$ has units of $\cm^{-2}$.
\end{thebibliography}

\end{document}